\documentstyle[12pt,epsfig]{article}
\begin{document}

\begin{flushright}
{\bf hep-ph/0301024} \\
{BIHEP-TH-2003-1} 
\end{flushright}

\vspace{0.5cm}

\begin{center}
{\large\bf Final-state Rescattering and SU(3) Symmetry Breaking \\
in $B\rightarrow DK$ and $B\rightarrow DK^*$ Decays}
\end{center}

\vspace{0.1cm}

\begin{center}
{\bf Zhi-zhong Xing}
\footnote{E-mail: xingzz@mail.ihep.ac.cn} \\
{\it CCAST (World Laboratory), P.O. Box 8730, Beijing 100080, China \\ 
and Institute of High Energy Physics, Chinese Academy of Sciences, \\
P.O. Box 918 (4), Beijing 100039, China}
\footnote{Mailing address}
\end{center}

\vspace{1cm}

\begin{abstract}
The first observation of $\bar{B}^0_d\rightarrow D^0\bar{K}^0$ and
$\bar{B}^0_d\rightarrow D^0\bar{K}^{*0}$ transitions by the Belle 
Collaboration allows us to do a complete isospin analysis of 
$B\rightarrow DK^{(*)}$ decay modes. We find that 
their respective isospin phase shifts are very likely to lie in
the ranges $37^\circ \leq (\phi_1 -\phi_0)_{DK} \leq 63^\circ$ 
(or around $50^\circ$) and 
$25^\circ \leq (\phi_1 -\phi_0)_{DK^*} \leq 50^\circ$ 
(or around $35^\circ$), although the possibility
$(\phi_1 -\phi_0)_{DK} = (\phi_1 -\phi_0)_{DK^*} = 0^\circ$ cannot
be ruled out at present. Thus significant final-state rescattering 
effects are possible to exist in such exclusive 
$|\Delta B| = |\Delta C| = |\Delta S| =1$ processes. 
We determine the spectator and color-suppressed 
spectator quark-diagram amplitudes of $B\rightarrow DK$ and 
$B\rightarrow DK^*$ decays, and compare them with the corresponding 
quark-diagram amplitudes of $B\rightarrow D\pi$ and $B\rightarrow D\rho$ 
decays. The effects of SU(3) flavor symmetry breaking are in most
cases understandable in the factorization approximation, which 
works for the individual isospin amplitudes. Very instructive predictions 
are also obtained for the branching fractions of rare
$\bar{B}^0_d \rightarrow \bar{D}^0 \bar{K}^{(*)0}$, 
$B^-_u \rightarrow \bar{D}^0 K^{(*)-}$ and
$B^-_u \rightarrow D^- \bar{K}^{(*)0}$ transitions.
\end{abstract}

\newpage

\section{Introduction}

The major goal of $B$-meson factories is to test the 
Kobayashi-Maskawa mechanism of CP violation within the standard model 
and to detect possible new sources of CP violation beyond the
standard model. So far the CP-violating asymmetry in
$B^0_d$ vs $\bar{B}^0_d \rightarrow J/\psi K_{\rm S}$ decays has been
unambiguously measured at KEK and SLAC \cite{2B}, 
and the experimental result is compatible very well with the standard-model 
expectation. Further experiments will provide much more data on 
CP violation in other decay modes of $B$ mesons, from which one may
cross-check the consistency of the Kobayashi-Maskawa picture and probe
possible new physics.

Two-body nonleptonic decays of the type 
$\bar{B}^0_d \rightarrow X^+ Y^-$, 
$\bar{B}^0_d \rightarrow X^0 \bar{Y}^0$, 
$B^-_u \rightarrow X^0 Y^-$ 
and their CP-conjugate modes, which occur only via the tree-level quark 
diagrams and can be related to one another via the isospin triangles,  
have been of great interest in $B$ physics for a stringent test of 
the factorization hypothesis, a quantitative analysis of final-state 
interactions \cite{Stech}, 
and a clean determination of CP-violating phases \cite{Fleischer}. The
typical examples are $X = (D, D^*)$ and $Y=(\pi, \rho)$ as well
as $X = (D, D^*)$ and $Y = (K, K^*)$, associated respectively with the 
weak phases $(2\beta + \gamma)$ \cite{Xing95} and $\gamma$ \cite{Gronau}. 
Once the decay rates of $\bar{B}^0_d \rightarrow X^+ Y^-$, 
$\bar{B}^0_d \rightarrow X^0 \bar{Y}^0$ and $B^-_u \rightarrow X^0 Y^-$
are all measured to an acceptable degree of accuracy, one may construct
the isospin relations of their transition amplitudes and determine
the relevant strong and weak phases. Among three channels under
discussion, the neutral one $\bar{B}^0_d \rightarrow X^0 \bar{Y}^0$ is
color-suppressed and has the smallest branching ratio. Hence 
it is most difficult to measure $\bar{B}^0_d \rightarrow X^0 \bar{Y}^0$
in practice, even at $B$-meson factories. Indeed 
$\bar{B}^0_d \rightarrow D^0\pi^0$ and $D^{*0}\pi^0$ decays were not  
observed until 2001 \cite{CLEO}. The isospin analysis of three 
$B \rightarrow D^{(*)}\pi$ modes \cite{Xing01} indicates that they 
involve significant final-state rescattering effects.

Recently the Belle Collaboration \cite{Belle} has reported the
first observation of $\bar{B}^0_d\rightarrow D^0\bar{K}^0$ and
$\bar{B}^0_d\rightarrow D^0\bar{K}^{*0}$ decays. Their branching
fractions are found to be 
\begin{eqnarray}
{\cal B}^{DK}_{00} & = & \left (5.0^{+1.3}_{-1.2} \pm 0.6 \right ) 
\times 10^{-5} \; ,
\nonumber \\
{\cal B}^{DK^*}_{00} & = & \left (4.8^{+1.1}_{-1.0} \pm 0.5 \right ) 
\times 10^{-5} \; ,
\end{eqnarray}
respectively. In comparison, the branching fractions of 
$\bar{B}^0_d\rightarrow D^+K^{(*)-}$ and 
$B^-_u\rightarrow D^0 K^{(*)-}$ decays are \cite{PDG}
\begin{eqnarray}
{\cal B}^{DK}_{+-} & = & \left (2.0 \pm 0.6 \right ) \times 10^{-4} \; ,
\nonumber \\
{\cal B}^{DK^*}_{+-} & = & \left (3.7 \pm 1.8 \right ) 
\times 10^{-4} \; ;
\end{eqnarray}
and
\begin{eqnarray}
{\cal B}^{DK}_{0-} & = & \left (3.7 \pm 0.6 \right ) \times 10^{-4} \; ,
\nonumber \\
{\cal B}^{DK^*}_{0-} & = & \left (6.1 \pm 2.3 \right ) 
\times 10^{-4} \; .
\end{eqnarray}
We see that 
${\cal B}^{DK}_{+-} \sim {\cal B}^{DK}_{0-} > {\cal B}^{DK}_{00}$
and 
${\cal B}^{DK^*}_{+-} \sim {\cal B}^{DK^*}_{0-} > {\cal B}^{DK^*}_{00}$
do hold, as naively expected. With the help of the new experimental 
data, one is now able to analyze the isospin relations for the
amplitudes of $B\rightarrow DK^{(*)}$ decays in a more complete 
way than before (see, e.g., Refs. \cite{DD} and \cite{Xing98}). Then it
becomes possible to examine whether final-state interactions are 
significant or not in such exclusive 
$|\Delta B| = |\Delta C| = |\Delta S| =1$ transitions. 

The purpose of this paper is four-fold. First, we make use of current
experimental data to determine the isospin amplitudes and their relative 
phase for $B\rightarrow DK^{(*)}$ transitions. We find that the 
isospin phase shifts in $B\rightarrow DK$ and $B\rightarrow DK^*$ decays
are very likely to lie in the ranges 
$37^\circ \leq (\phi_1 -\phi_0)_{DK} \leq 63^\circ$ (or around
$50^\circ$) and $25^\circ \leq (\phi_1 -\phi_0)_{DK^*} \leq 50^\circ$ 
(or around $35^\circ$), although the possibility
$(\phi_1 -\phi_0)_{DK} = (\phi_1 -\phi_0)_{DK^*} = 0^\circ$ cannot
be ruled out at the moment. Hence significant final-state rescattering 
effects are possible to exist in these interesting
$|\Delta B| = |\Delta C| = |\Delta S| =1$ processes. 
Second, we carry out a quark-diagram analysis 
of $B\rightarrow DK^{(*)}$ transitions and determine the amplitudes
of their spectator and color-suppressed spectator diagrams. Our result
is model-independent, thus it can be used to test the predictions
from specific models of hadronic matrix elements. Third, we compare 
the quark-diagram amplitudes of $B\rightarrow DK$ and
$B\rightarrow DK^*$ with those of $B\rightarrow D\pi$ and
$B\rightarrow D\rho$ to examine the SU(3) flavor symmetry. We find 
that the effects of SU(3) symmetry breaking are in most cases 
understandable in the factorization approximation, which works
well for the individual isospin amplitudes. Finally, we present very 
instructive predictions for the branching fractions of rare
$\bar{B}^0_d \rightarrow \bar{D}^0 \bar{K}^{(*)0}$, 
$B^-_u \rightarrow \bar{D}^0 K^{(*)-}$ and
$B^-_u \rightarrow D^- \bar{K}^{(*)0}$ decays. 

\section{Isospin Analysis}

Let us begin with an isospin analysis of 
$\bar{B}^0_d \rightarrow D^+ K^-$, $\bar{B}^0_d \rightarrow D^0 \bar{K}^0$
and $B^-_u \rightarrow D^0 K^-$ decays. As $D$ and $K$ are isospin 1/2
mesons, a final state $DK$ may be in either $I=1$ or $I=0$ configuration.
To be specific, $D^0K^-$ is a pure $I=1$ state and involves a single
rescattering phase. In contrast, $D^+K^-$ and $D^0\bar{K}^0$ are
combinations of $I=1$ and $I=0$ states, which mix under rescattering. 
The amplitudes of three decay modes can therefore be expressed as
\begin{eqnarray}
A^{DK}_{+-} & = & \frac{1}{2} A^{DK}_1 + \frac{1}{2} A^{DK}_0 \; ,
\nonumber \\
A^{DK}_{00} & = & \frac{1}{2} A^{DK}_1 - \frac{1}{2} A^{DK}_0 \; ,
\nonumber \\
A^{DK}_{0-} & = &  A^{DK}_1
\end{eqnarray}
in terms of two isospin amplitudes $A^{DK}_1$ and $A^{DK}_0$. The weak
phases of $A^{DK}_1$ and $A^{DK}_0$ are identical to 
$\arg (V_{cb}V^*_{us})$ \cite{DD}, where $V_{cb}$ and $V_{us}$ are
the Cabibbo-Kobayashi-Maskawa (CKM) matrix elements. The strong phases
of $A^{DK}_1$ and $A^{DK}_0$, denoted respectively as $\phi_1$ and
$\phi_0$, are in general different from each other.
The branching fractions of $\bar{B}^0_d \rightarrow D^+ K^-$, 
$\bar{B}^0_d \rightarrow D^0 \bar{K}^0$ and $B^-_u \rightarrow D^0 K^-$ 
transitions read 
\begin{eqnarray}
{\cal B}^{DK}_{+-} & = & \frac{p^{~}_{DK}}{8\pi M^2_B} 
\left |A^{DK}_{+-} \right |^2 \tau^{~}_0 \; ,
\nonumber \\
{\cal B}^{DK}_{00} & = & \frac{p^{~}_{DK}}{8\pi M^2_B} 
\left |A^{DK}_{00} \right |^2 \tau^{~}_0 \; ,
\nonumber \\
{\cal B}^{DK}_{0-} & = & \frac{p^{~}_{DK}}{8\pi M^2_B} 
\left |A^{DK}_{0-} \right |^2 \tau^{~}_{\pm} \; ,
\end{eqnarray}
where $\tau^{~}_0$ (or $\tau^{~}_{\pm}$) is the lifetime of $B^0_d$ or
$\bar{B}^0_d$ (or $B^\pm_u$), and
\begin{equation}
p^{~}_{DK} \; =\; \frac{1}{2M_B} \sqrt{\left [ M^2_B - 
(M_D + M_K)^2
\right ] \left [ M^2_B - (M_D - M_K)^2 \right ]} \; 
\end{equation}
is the c.m. momentum of $D$ and $K$ mesons. Note that the mass 
difference between neutral and charged $B$, $D$ or $K$ mesons is tiny 
and has been neglected in Eqs. (5) and (6). From Eqs. (4) and (5),
we obtain
\begin{eqnarray}
\left |A^{DK}_1 \right | & = & 
2M_B \sqrt{\frac{2\pi \kappa {\cal B}^{DK}_{0-}}
{\tau^{~}_0 p^{~}_{DK}}} \;\; ,
\nonumber \\
\left |A^{DK}_0 \right | & = & 
2M_B \sqrt{\frac{2\pi \left [ 2 \left (
{\cal B}^{DK}_{+-} + {\cal B}^{DK}_{00} \right ) - 
\kappa {\cal B}^{DK}_{0-} \right ]}{\tau^{~}_0 p^{~}_{DK}}} \;\; 
\end{eqnarray}
with $\kappa \equiv \tau^{~}_0/\tau^{~}_{\pm} \approx 0.92$ \cite{PDG};
and
\begin{equation}
\cos (\phi_1 - \phi_0)^{~}_{DK} \; =\; \frac{{\cal B}^{DK}_{+-} -
{\cal B}^{DK}_{00}}{\sqrt{ \kappa {\cal B}^{DK}_{0-} 
\left [ 2 \left ({\cal B}^{DK}_{+-} + {\cal B}^{DK}_{00} \right ) -
\kappa {\cal B}^{DK}_{0-} \right ]}} \; .
\end{equation}
If final-state interactions were insignificant, one would have 
$\cos (\phi_1 - \phi_0)^{~}_{DK} \approx 1$. In this case, the
naive factorization approximation could be applied to the 
{\it overall} amplitudes of three decay modes under discussion.

To illustrate, we take the experimental values of 
${\cal B}^{DK}_{+-}$, ${\cal B}^{DK}_{00}$ and ${\cal B}^{DK}_{0-}$
in Eqs. (1), (2) and (3) to calculate the isospin amplitudes 
$A^{DK}_1$ and $A^{DK}_0$. The values of other input 
parameters in Eqs. (7) and (8) can be found from Ref. \cite{PDG}. 
Our numerical results are shown in Fig. 1, from which we obtain
$1.93 \times 10^{-7} ~ {\rm GeV} \leq |A^{DK}_1| 
\leq 2.28 \times 10^{-7} ~ {\rm GeV}$,
$0.44 \times 10^{-7} ~ {\rm GeV} \leq |A^{DK}_0| 
\leq 2.19 \times 10^{-7} ~ {\rm GeV}$, and 
$0.41 \leq \cos(\phi_1 -\phi_0)_{DK} \leq 1$. Although the
possibility $(\phi_1 -\phi_0)_{DK} =0^\circ$ cannot be excluded
for the time being \cite{Rosner}, 
Fig. 1 indicates that this isospin phase shift is most likely to 
lie in the range $37^\circ \leq (\phi_1 -\phi_0)_{DK} \leq 63^\circ$.
In particular, the central values of $|A^{DK}_1|$,
$|A^{DK}_0|$ and $(\phi_1 -\phi_0)_{DK}$ read as 
\begin{eqnarray}
\left |A^{DK}_1 \right | & \approx & 2.1 \times 10^{-7} ~ 
{\rm GeV} \; , 
\nonumber \\
\left |A^{DK}_0 \right | & \approx & 1.4 \times 10^{-7} ~ 
{\rm GeV} \; , 
\end{eqnarray}
and
\begin{equation}
\left (\phi_1 - \phi_0 \right )^{~}_{DK} \; \approx \; 50^\circ \; .
\end{equation}
Our results imply that significant final-state rescattering effects are 
very likely to exist in $B\rightarrow DK$ decays. Hence the neutral mode
$\bar{B}^0_d \rightarrow D^0 \bar{K}^0$ may not be strongly suppressed,
even though $A^{DK}_1$ and $A^{DK}_0$ are comparable in magnitude.

As $B\rightarrow DK^*$ decays have the same isospin relations
as $B\rightarrow DK$ decays, one may carry out an analogous analysis
to determine the isospin amplitudes of the former 
($A^{DK^*}_1$ and $A^{DK^*}_0$) by use of the 
experimental values of ${\cal B}^{DK^*}_{+-}$, ${\cal B}^{DK^*}_{00}$ and 
${\cal B}^{DK^*}_{0-}$ in Eqs. (1), (2) and (3). Our 
results are presented in Fig. 2, from which
$2.17 \times 10^{-7} ~ {\rm GeV} \leq |A^{DK^*}_1| 
\leq 3.23 \times 10^{-7} ~ {\rm GeV}$,
$0.71 \times 10^{-7} ~ {\rm GeV} \leq |A^{DK^*}_0| 
\leq 3.42 \times 10^{-7} ~ {\rm GeV}$, and 
$0.58 \leq \cos(\phi_1 -\phi_0)_{DK^*} \leq 1$ are directly
obtainable. Note again that the possibility 
$(\phi_1 -\phi_0)_{DK^*} =0^\circ$ 
cannot be ruled out at the moment \cite{Rosner},
but Fig. 2 indicates that this isospin phase shift is most likely to 
lie in the range $25^\circ \leq (\phi_1 -\phi_0)_{DK^*} \leq 50^\circ$.
The central values of $|A^{DK^*}_1|$, $|A^{DK^*}_0|$ and 
$(\phi_1 -\phi_0)_{DK^*}$ are actually 
\begin{eqnarray}
\left |A^{DK^*}_1 \right | & \approx & 2.8 \times 10^{-7} ~ 
{\rm GeV} \; , 
\nonumber \\
\left |A^{DK^*}_0 \right | & \approx & 1.9 \times 10^{-7} ~ 
{\rm GeV} \; , 
\end{eqnarray}
and
\begin{equation}
\left (\phi_1 - \phi_0 \right )^{~}_{DK^*} \; \approx \; 35^\circ \; .
\end{equation}
It becomes clear that large final-state interactions are also likely to
exist in $B\rightarrow DK^*$ decays. 
Our results imply that the naive factorization
approximation should not be {\it directly} applied to both $B\rightarrow DK$ 
and $B\rightarrow DK^*$ transitions. A proper treatment of such exclusive
processes has to take account of final-state rescattering effects at the
hadron level.

\section{Quark-diagram analysis}

Now we turn to describe $B\rightarrow DK$ and $B\rightarrow DK^*$ decays
in the language of quark diagrams, which is sometimes more intuitive
and instructive than the isospin language. As shown in Fig. 3, these 
processes occur only via two tree-level quark diagrams: one is the 
spectator diagram and the other is the color-suppressed spectator diagram. 
After the CKM matrix elements ($V_{cb}$ and $V^*_{us}$) are factored out,
the remaining parts of the quark-diagram amplitudes in Fig. 3(a) and
Fig. 3(b) can be defined under isospin symmetry as $T_{DK}$ (or
$T_{DK^*}$) and $T'_{DK}$ (or $T'_{DK^*}$), respectively. The relations
between quark-diagram amplitudes and isospin amplitudes of 
$B\rightarrow DK$ transitions are
\begin{eqnarray}
A^{DK}_1 & = & V_{cb}V^*_{us} \left (T_{DK} + T'_{DK} \right )
e^{i\phi_1} \; ,
\nonumber \\
A^{DK}_0 & = & V_{cb}V^*_{us} \left (T_{DK} - T'_{DK} \right )
e^{i\phi_0} \; .
\end{eqnarray}
Without loss of generality, $T_{DK}$ and $T'_{DK}$ can be arranged
to be real and positive. In addition, $T_{DK} > T'_{DK}$ is expected
to hold, as the latter is color-suppressed. The explicit expressions 
of $T_{DK}$ and $T'_{DK}$ can straightforwardly be derived from 
Eqs. (4), (5) and (13). The result is
\begin{eqnarray}
T_{DK} & = & \frac{M_B}{|V_{cb}V^*_{us}|} \sqrt{\frac{2\pi}
{\tau^{~}_0 p^{~}_{DK}}} \left [\sqrt{\kappa {\cal B}^{DK}_{0-}} ~ +
\sqrt{2 \left ( {\cal B}^{DK}_{+-} + {\cal B}^{DK}_{00} \right ) -
\kappa {\cal B}^{DK}_{0-}} \right ] \; ,
\nonumber \\
T'_{DK} & = & \frac{M_B}{|V_{cb}V^*_{us}|} \sqrt{\frac{2\pi}
{\tau^{~}_0 p^{~}_{DK}}} \left [\sqrt{\kappa {\cal B}^{DK}_{0-}} ~ -
\sqrt{2 \left ( {\cal B}^{DK}_{+-} + {\cal B}^{DK}_{00} \right ) -
\kappa {\cal B}^{DK}_{0-}} \right ] \; .
\end{eqnarray}
Of course, it is easy to obtain the similar expressions for $T_{DK^*}$ 
and $T'_{DK^*}$, the quark-diagram amplitudes of $B\rightarrow DK^*$ 
decay modes.

It is worth emphasizing that Eq. (14) allows us to determine $T_{DK}$ 
(or $T_{DK^*}$) and $T'_{DK}$ (or $T'_{DK^*}$) in a model-independent way. 
On the other hand, these quark-diagram amplitudes can be
evaluated using the factorization approximation in specific models of 
hadronic matrix elements \cite{Stech}. Thus the validity of the 
factorization hypothesis for $B\rightarrow DK$ and $B\rightarrow DK^*$ 
decays is experimentally testable 
\footnote{Detailed reanalyses of $B\rightarrow DK$ and 
$B\rightarrow DK^*$ transitions in the factorization approximation
and in various form-factor models will be presented in a more
comprehensive article \cite{X}. However, some concise discussions can 
be found in section 4 to phenomenologically understand the effects of 
SU(3) symmetry breaking in the factorization approximation.}.

For illustration, we calculate $(T_{DK}, T'_{DK})$ and 
$(T_{DK^*}, T'_{DK^*})$ numerically by use of the central values of
$({\cal B}^{DK}_{+-}, {\cal B}^{DK}_{00}, {\cal B}^{DK}_{0-})$
and $({\cal B}^{DK^*}_{+-}, {\cal B}^{DK^*}_{00}, {\cal B}^{DK^*}_{0-})$ 
given in Eqs. (1), (2) and (3). The relevant CKM matrix elements are
taken as $|V_{cb}| = 0.041$ and $|V_{us}| = 0.222$ \cite{PDG}. 
We arrive at
\begin{eqnarray}
T_{DK} & \approx & 1.9 \times 10^{-5} ~ {\rm GeV} \; ,
\nonumber \\
T'_{DK} & \approx & 3.7 \times 10^{-6} ~ {\rm GeV} \; ;
\end{eqnarray}
and 
\footnote{Note that the polarization effect of $K^*$ in the final 
states of $B\rightarrow DK^*$ decays has been included in 
$T_{DK^*}$ and $T'_{DK^*}$. This point will become transparent
in Eq. (24), where $T_{DK^*}$ and $T'_{DK^*}$ are calculated in the 
factorization approximation.}
\begin{eqnarray}
T_{DK^*} & \approx & 2.6 \times 10^{-5} ~ {\rm GeV} \; ,
\nonumber \\
T'_{DK^*} & \approx & 4.5 \times 10^{-6} ~ {\rm GeV} \; .
\end{eqnarray}
One can see that $T'_{DK}/T_{DK} \sim T'_{DK^*}/T_{DK^*} \sim 0.2$
holds. Such an instructive and model-independent result is essentially 
consistent with the naive expectation for $T'_{DK}/T_{DK}$ and 
$T'_{DK^*}/T_{DK^*}$ 
($\sim a^{\rm eff}_2/a^{\rm eff}_1 \approx 0.25$ \cite{Stech}) in the 
factorization approximation.

\section{SU(3) symmetry breaking}

We have pointed out that $B\rightarrow DK$ and $B\rightarrow D\pi$
transitions belong to the same category of exclusive $B$ decays
with $|\Delta B| = |\Delta C| =1$. They can be related
to each other under SU(3) flavor symmetry \cite{Gronau2}. To examine
how good (or bad) this SU(3) symmetry is, let us write out the 
amplitudes of $\bar{B}^0_d \rightarrow D^+\pi^-$, 
$\bar{B}^0_d \rightarrow D^0\pi^0$ and $B^-_u \rightarrow D^0\pi^-$ 
decays in terms of two isospin amplitudes $A^{D\pi}_{3/2}$ and
$A^{D\pi}_{1/2}$:
\begin{eqnarray}
A^{D\pi}_{+-} & = & \frac{1}{\sqrt{3}} A^{D\pi}_{3/2} + 
\frac{\sqrt{2}}{\sqrt{3}} A^{D\pi}_{1/2} \; ,
\nonumber \\
A^{D\pi}_{00} & = & \frac{\sqrt{2}}{\sqrt{3}} A^{D\pi}_{3/2} - 
\frac{1}{\sqrt{3}} A^{D\pi}_{1/2} \; ,
\nonumber \\
A^{D\pi}_{0-} & = & \sqrt{3} A^{D\pi}_{3/2} \; .
\end{eqnarray}
Note that these three decay modes can in general occur via three 
topologically distinct tree-level quark diagrams \cite{Xing98}: the 
spectator diagram 
($\propto V_{cb}V^*_{ud} T_{D\pi}$) similar to Fig. 3(a), 
the color-suppressed diagram 
($\propto V_{cb}V^*_{ud} T'_{D\pi}$) similar to Fig. 3(b), and the 
$W$-exchange diagram ($\propto V_{cb}V^*_{ud} T''_{D\pi}$). In
comparison with $T_{D\pi}$ and $T'_{D\pi}$, $T''_{D\pi}$ is 
expected to have strong form-factor suppression \cite{Xing96}.
It is therefore safe, at least to leading order, to neglect the
contribution of $T''_{D\pi}$ to the overall amplitudes of 
$B\rightarrow D\pi$ decays. In this approximation, we have
\begin{eqnarray}
A^{D\pi}_{3/2} & = & \frac{1}{\sqrt{3}} V_{cb}V^*_{ud} 
\left (T_{D\pi} + T'_{D\pi} \right ) e^{i\phi_{3/2}} \; ,
\nonumber \\
A^{D\pi}_{1/2} & = & \frac{1}{\sqrt{6}} V_{cb}V^*_{ud} 
\left (2 T_{D\pi} - T'_{D\pi} \right ) e^{i\phi_{1/2}} \; ,
\end{eqnarray}
where $\phi_{3/2}$ and $\phi_{1/2}$ are the strong phases of $I=3/2$ 
and $I=1/2$ isospin channels. If SU(3) were a perfect symmetry,
one would get $T_{D\pi} = T_{DK}$ and $T'_{D\pi} = T'_{DK}$.

Similar to Eq. (14), the explicit expressions of $T_{D\pi}$ and 
$T'_{D\pi}$ can be obtained in terms of the branching fractions
of $\bar{B}^0_d \rightarrow D^+\pi^-$ (${\cal B}^{D\pi}_{+-}$), 
$\bar{B}^0_d \rightarrow D^0\pi^0$ (${\cal B}^{D\pi}_{00}$)
and $B^-_u \rightarrow D^0\pi^-$ (${\cal B}^{D\pi}_{0-}$). Then
we arrive at the ratios $T_{DK}/T_{D\pi}$ and $T'_{DK}/T'_{D\pi}$ 
as follows:
\begin{eqnarray}
\frac{T_{DK}}{T_{D\pi}} & = & \frac{3}{2} \left | \frac{V_{ud}}
{V_{us}} \right | \sqrt{\frac{p^{~}_{D\pi}}{p^{~}_{DK}}} ~
\frac{\sqrt{\kappa {\cal B}^{DK}_{0-}} ~ +
\sqrt{2 \left ( {\cal B}^{DK}_{+-} + {\cal B}^{DK}_{00} \right ) -
\kappa {\cal B}^{DK}_{0-}}}
{\sqrt{\kappa {\cal B}^{D\pi}_{0-}} ~ +
\sqrt{6 \left ( {\cal B}^{D\pi}_{+-} + {\cal B}^{D\pi}_{00} \right ) -
2 \kappa {\cal B}^{D\pi}_{0-}}} \; ,
\nonumber \\
\frac{T'_{DK}}{T'_{D\pi}} & = & \frac{3}{2} \left | \frac{V_{ud}}
{V_{us}} \right | \sqrt{\frac{p^{~}_{D\pi}}{p^{~}_{DK}}} ~
\frac{\sqrt{\kappa {\cal B}^{DK}_{0-}} ~ -
\sqrt{2 \left ( {\cal B}^{DK}_{+-} + {\cal B}^{DK}_{00} \right ) -
\kappa {\cal B}^{DK}_{0-}}}
{2 \sqrt{\kappa {\cal B}^{D\pi}_{0-}} ~ -
\sqrt{6 \left ( {\cal B}^{D\pi}_{+-} + {\cal B}^{D\pi}_{00} \right ) -
2 \kappa {\cal B}^{D\pi}_{0-}}} \; ,
\end{eqnarray}
where 
\begin{equation}
p^{~}_{D\pi} \; =\; \frac{1}{2M_B} \sqrt{\left [ M^2_B - 
(M_D + M_\pi)^2
\right ] \left [ M^2_B - (M_D - M_\pi)^2 \right ]} \; 
\end{equation}
is the c.m. momentum of $D$ and $\pi$ mesons. Typically taking the
central values of ${\cal B}^{D\pi}_{+-}$, ${\cal B}^{D\pi}_{00}$
and ${\cal B}^{D\pi}_{0-}$ reported in Ref. \cite{PDG}, we obtain
\begin{eqnarray}
\frac{T_{DK}}{T_{D\pi}} & \approx & 1.21 \; , 
\nonumber \\
\frac{T'_{DK}}{T'_{D\pi}} & \approx & 0.97 \; .
\end{eqnarray}
This model-independent result can be confronted with the result
achieved from the naive factorization approximation:
\begin{eqnarray}
\frac{T_{DK}}{T_{D\pi}} & = & \frac{f_K}{f_\pi} \cdot
\frac{F^{BD}_0(M^2_K)}{F^{BD}_0(M^2_\pi)}
\nonumber \\ 
& \approx & 1.22 \; ,
\nonumber \\
\frac{T'_{DK}}{T'_{D\pi}} & = & \frac{M^2_B - M^2_K}{M^2_B - M^2_\pi}
\cdot \frac{F^{BK}_0(M^2_D)}{F^{B\pi}_0(M^2_D)} 
\nonumber \\
& \approx & 1.09 \; ,
\end{eqnarray}
where $f_\pi = 130.7$ MeV, $f_K = 159.8$ MeV \cite{PDG}, 
$F^{B\pi}_0(0) = 0.28$ and $F^{BK}_0(0) = 0.31$ \cite{Stech}
have been used. We observe that Eqs. (21) and (22) are 
consistent with each other. It is suggestive that the factorization
hypothesis may work well for the individual isospin amplitudes
of $B\rightarrow DK$ and $B\rightarrow D\pi$ transitions.

One can analogously compare between the quark-diagram amplitudes
of $B\rightarrow DK^*$ and $B\rightarrow D\rho$ decays, i.e.,
$T_{DK^*}$ (or $T'_{DK^*}$) and $T_{D\rho}$ (or $T'_{D\rho}$),
to estimate the size of SU(3) flavor symmetry breaking. Typically 
taking the central value of ${\cal B}^{D\rho}_{00}$ reported by the
Belle Collaboration \cite{Belle2} and those of 
${\cal B}^{D\rho}_{+-}$ and ${\cal B}^{D\rho}_{0-}$ reported in
Ref. \cite{PDG}, we obtain
\begin{eqnarray}
\frac{T_{DK^*}}{T_{D\rho}} & \approx & 1.00 \; , 
\nonumber \\
\frac{T'_{DK^*}}{T'_{D\rho}} & \approx & 0.70 \; .
\end{eqnarray}
In comparison, the factorization approximation yields
\begin{eqnarray}
\frac{T_{DK^*}}{T_{D\rho}} & = & \frac{M_{K^*}}{M_\rho} \cdot
\frac{f_{K^*}}{f_\rho} \cdot \frac{F^{BD}_1(M^2_{K^*})}{F^{BD}_1(M^2_\rho)}
\left | \frac{\epsilon^{~}_{K^*} \cdot p^{~}_B}
{\epsilon^{~}_\rho \cdot p^{~}_B} \right | 
\nonumber \\
& = & \frac{p^{~}_{DK^*}}{p^{~}_{D\rho}} \cdot
\frac{f_{K^*}}{f_\rho} \cdot \frac{F^{BD}_1(M^2_{K^*})}{F^{BD}_1(M^2_\rho)}
\nonumber \\
& \approx & 1.02 \; ,
\nonumber \\
\frac{T'_{DK^*}}{T'_{D\rho}} & = & \frac{M_{K^*}}{M_\rho} \cdot
\frac{A^{BK^*}_0(M^2_D)}{A^{B\rho}_0(M^2_D)} 
\left | \frac{\epsilon^{~}_{K^*} \cdot p^{~}_D}
{\epsilon^{~}_\rho \cdot p^{~}_D} \right | 
\nonumber \\
& = & \frac{p^{~}_{DK^*}}{p^{~}_{D\rho}} \cdot
\frac{M_B - 2 \sqrt{M^2_{K^*} + p^2_{DK^*}}}
{M_B - 2 \sqrt{M^2_\rho + p^2_{D\rho}}} \cdot
\frac{A^{BK^*}_0(M^2_D)}{A^{B\rho}_0(M^2_D)} 
\nonumber \\
& \approx & 1.16 \; ,
\end{eqnarray}
where $f_{K^*} = 214$ MeV, $f_\rho = 210$ MeV, $A^{BK^*}_0(0) = 0.47$
and $A^{B\rho}_0(0) = 0.37$ \cite{Stech} have typically been used.
We see that the results for $T_{DK^*}/T_{D\rho}$ in Eqs. (23) and
(24) are in good agreement, but there is a remarkable discrepancy
between the model-dependent and model-independent results for 
$T'_{DK^*}/T'_{D\rho}$. 

\section{Further discussions}

The present experimental data on $B\rightarrow DK$ and
$B\rightarrow DK^*$ decays in Eqs. (1), (2) and (3) allow us to 
determine their isospin and quark-diagram amplitudes in a 
quantitatively meaningful way. It should be noted, however, that
the accuracy of ${\cal B}^{DK^*}_{+-}$ and ${\cal B}^{DK^*}_{0-}$ 
is rather poor. Hence the result $T'_{DK^*}/T'_{D\rho} \approx 0.7$ 
in Eq. (23) is most likely a signal of poor accuracy of current data,
instead of a signal of significant SU(3) symmetry breaking. As one
can see from Eq. (19), $T'_{DK^*}/T'_{D\rho}$ may be
more sensitive to the uncertainties of ${\cal B}^{DK^*}_{+-}$ and 
${\cal B}^{DK^*}_{0-}$ than $T_{DK^*}/T_{D\rho}$, because the
former involves large cancellations in both its numerator and 
denominator. We therefore argue that the existing discrepancy between
the model-dependent and model-independent results of 
$T'_{DK^*}/T'_{D\rho}$ could essentially disappear, when more
precise measurements of $B\rightarrow DK^*$ decays are available.

It is worth remarking that only the central values of relevant
branching fractions of $B\rightarrow DK^{(*)}$ decays have been
taken into account in our numerical estimates of the quark-diagram
amplitudes and SU(3) symmetry breaking effects. In addition, current 
experimental error bars associated with a few of those branching 
fractions remain too large to allow for a firm quantitative
conclusion about the size and phase of final-state rescattering 
in every $|\Delta B| = |\Delta C| = |\Delta S| =1$ mode. The same 
point has been emphasized by Chiang and Rosner in Ref. \cite{Rosner}.
Hopefully, more accurate data will soon be available at $B$-meson
factories.

In general, the $B\rightarrow DK^{(*)}$ decays include
not only $\bar{B}^0_d \rightarrow D^+ K^{(*)-}$, 
$\bar{B}^0_d \rightarrow D^0 \bar{K}^{(*)0}$ and
$B^-_u \rightarrow D^0 K^{(*)-}$ but also
$\bar{B}^0_d \rightarrow \bar{D}^0 \bar{K}^{(*)0}$, 
$B^-_u \rightarrow \bar{D}^0 K^{(*)-}$ and
$B^-_u \rightarrow D^- \bar{K}^{(*)0}$. So far only the former
have been measured at $B$-meson factories. The latter have 
lower branching fractions, because of the stronger CKM suppression: 
$|V_{ub}V^*_{cs}|^2/|V_{cb}V^*_{us}|^2 \approx 0.15$ \cite{PDG}.
Note that the isospin relations among three amplitudes of 
$\bar{B}^0_d \rightarrow \bar{D}^0 \bar{K}^{(*)0}$, 
$B^-_u \rightarrow \bar{D}^0 K^{(*)-}$ and
$B^-_u \rightarrow D^- \bar{K}^{(*)0}$ transitions are quite
similar to those given in Eq. (4) \cite{DD}. In particular,
the unknown branching fractions of 
$\bar{B}^0_d \rightarrow \bar{D}^0 \bar{K}^0$ 
($\tilde{\cal B}^{DK}_{00}$) and
$\bar{B}^0_d \rightarrow \bar{D}^0 \bar{K}^{*0}$ 
($\tilde{\cal B}^{DK^*}_{00}$) can be predicted from
the measured branching fractions of 
$B^-_u \rightarrow D^0 K^-$ and $B^-_u \rightarrow D^0 K^{*-}$.
The results are
\begin{eqnarray}
\tilde{\cal B}^{DK}_{00} & = & \kappa \left |\frac{V_{ub}V^*_{cs}}
{V_{cb}V^*_{us}} \right |^2 \left (\frac{T'_{DK}}{T_{DK} + T'_{DK}}
\right )^2 {\cal B}^{DK}_{0-} 
\nonumber \\
& \approx & 1.4 \times 10^{-6} \; ,
\nonumber \\
\tilde{\cal B}^{DK^*}_{00} & = & \kappa \left |\frac{V_{ub}V^*_{cs}}
{V_{cb}V^*_{us}} \right |^2 \left (\frac{T'_{DK^*}}{T_{DK^*} + T'_{DK^*}}
\right )^2 {\cal B}^{DK^*}_{0-} 
\nonumber \\
& \approx & 1.8 \times 10^{-6} \; ,
\end{eqnarray}
well below the experimental upper bounds reported by the Belle 
Collaboration \cite{Belle}. If the phase shift between
$I=1$ and $I=0$ isospin channels in 
$B^-_u \rightarrow \bar{D}^0 K^{(*)-}$ and 
$B^-_u \rightarrow D^- \bar{K}^{(*)0}$ transitions is assumed to 
equal the corresponding phase shift between $I=1$ and $I=0$ 
isospin channels in $\bar{B}^0_d \rightarrow D^+ K^{(*)-}$ and
$\bar{B}^0_d \rightarrow D^0 \bar{K}^{(*)0}$ decays \cite{DD,Xing98b}
\footnote{Such an assumption has been questioned in 
Ref. \cite{Gronau3}, but there have not been strong theoretical 
arguments to disprove it. The measurements 
of $B^-_u \rightarrow \bar{D}^0 K^{(*)-}$ and 
$B^-_u \rightarrow D^- \bar{K}^{(*)0}$ decays in the future at 
$B$-meson factories will allow us to clarify the present ambiguity 
associated with their isospin phases.},
then instructive predictions can be obtained for the 
branching fractions of 
$B^-_u \rightarrow \bar{D}^0 K^{(*)-}$ ($\tilde{\cal B}^{DK}_{0-}$
or $\tilde{\cal B}^{DK^*}_{0-}$) and
$B^-_u \rightarrow D^- \bar{K}^{(*)0}$ 
($\tilde{B}^{DK}_{-0}$ or $\tilde{B}^{DK^*}_{-0}$):
\begin{eqnarray}
\tilde{\cal B}^{DK}_{0-} & = & \frac{1}{\kappa} 
\tilde{\cal B}^{DK}_{00} \cos^2\frac{(\phi_1 -\phi_0)^{~}_{DK}}{2} 
\nonumber \\
& \approx & 1.2 \times 10^{-6} \; ,
\nonumber \\
\tilde{\cal B}^{DK^*}_{0-} & = & \frac{1}{\kappa}
\tilde{\cal B}^{DK^*}_{00} \cos^2\frac{(\phi_1 -\phi_0)^{~}_{DK^*}}{2} 
\nonumber \\
& \approx & 1.8 \times 10^{-6} \; ;
\end{eqnarray}
and
\begin{eqnarray}
\tilde{\cal B}^{DK}_{-0} & = & \frac{1}{\kappa} 
\tilde{\cal B}^{DK}_{00} \sin^2\frac{(\phi_1 -\phi_0)^{~}_{DK}}{2} 
\nonumber \\
& \approx & 2.7 \times 10^{-7} \; ,
\nonumber \\
\tilde{\cal B}^{DK^*}_{-0} & = & \frac{1}{\kappa}
\tilde{\cal B}^{DK^*}_{00} \sin^2\frac{(\phi_1 -\phi_0)^{~}_{DK^*}}{2} 
\nonumber \\
& \approx & 1.8 \times 10^{-7} \; .
\end{eqnarray}
In arriving at these model-independent results, we have neglected 
small contributions of the annihilation-type quark diagrams to
$B^-_u \rightarrow \bar{D}^0 K^{(*)-}$ and 
$B^-_u \rightarrow D^- \bar{K}^{(*)0}$ decays. Although both
charged modes have quite small branching fractions, they are
worth being searched for at $B$-meson factories. The reason
is simply that the transitions under discussion are very important
for a relatively clean determination of the weak angle $\gamma$
of the well-known CKM unitarity triangle \cite{Gronau}.

In summary, we have for the first time determined the isospin 
amplitudes and their 
relative phase for $B\rightarrow DK^{(*)}$ transitions by use
of current experimental data. It is found that the isospin phase 
shifts in $B\rightarrow DK$ and $B\rightarrow DK^*$ decays are
very likely to lie in the ranges 
$37^\circ \leq (\phi_1 -\phi_0)_{DK} \leq 63^\circ$ (or around
$50^\circ$) and 
$25^\circ \leq (\phi_1 -\phi_0)_{DK^*} \leq  50^\circ$ 
(or around $35^\circ$), although the possibility 
$(\phi_1 -\phi_0)_{DK} = (\phi_1 -\phi_0)_{DK^*} = 0^\circ$
cannot be excluded at present. Thus significant final-state 
rescattering effects are possible to exist in such exclusive 
$|\Delta B| = |\Delta C| = |\Delta S| =1$ processes.
We have also analyzed $B\rightarrow DK^{(*)}$ transitions in 
the language of quark diagrams, and determined the amplitudes
of their spectator and color-suppressed spectator diagrams. We
stress that our results are model-independent. The SU(3)
flavor symmetry is examined by comparing the quark-diagram 
amplitudes of $B\rightarrow DK$ and $B\rightarrow DK^*$ decays 
with those of $B\rightarrow D\pi$ and $B\rightarrow D\rho$ decays.
We find that the effects of SU(3) symmetry breaking are 
in most cases understandable in the factorization approximation,
which works for the individual isospin amplitudes.
Finally, very instructive predictions for the branching fractions of
$\bar{B}^0_d \rightarrow \bar{D}^0 \bar{K}^{(*)0}$, 
$B^-_u \rightarrow \bar{D}^0 K^{(*)-}$ and
$B^-_u \rightarrow D^- \bar{K}^{(*)0}$ transitions have been 
presented. We expect that more precise experimental data from 
$B$-meson factories will help us to gain a deeper understanding 
of the dynamics of $B\rightarrow DK^{(*)}$ decays and to probe 
the signal of CP violation in them.
 
\vspace{0.5cm}

I am  indebted to T. Browder for calling my attention
to the recent Belle measurement of the
$\bar{B}^0_d\rightarrow D^0\rho^0$ decay mode \cite{Belle2}.
This work was supported in part by National Natural 
Science Foundation of China.

\newpage

\begin{figure}[t]
\vspace{5cm}
\epsfig{file=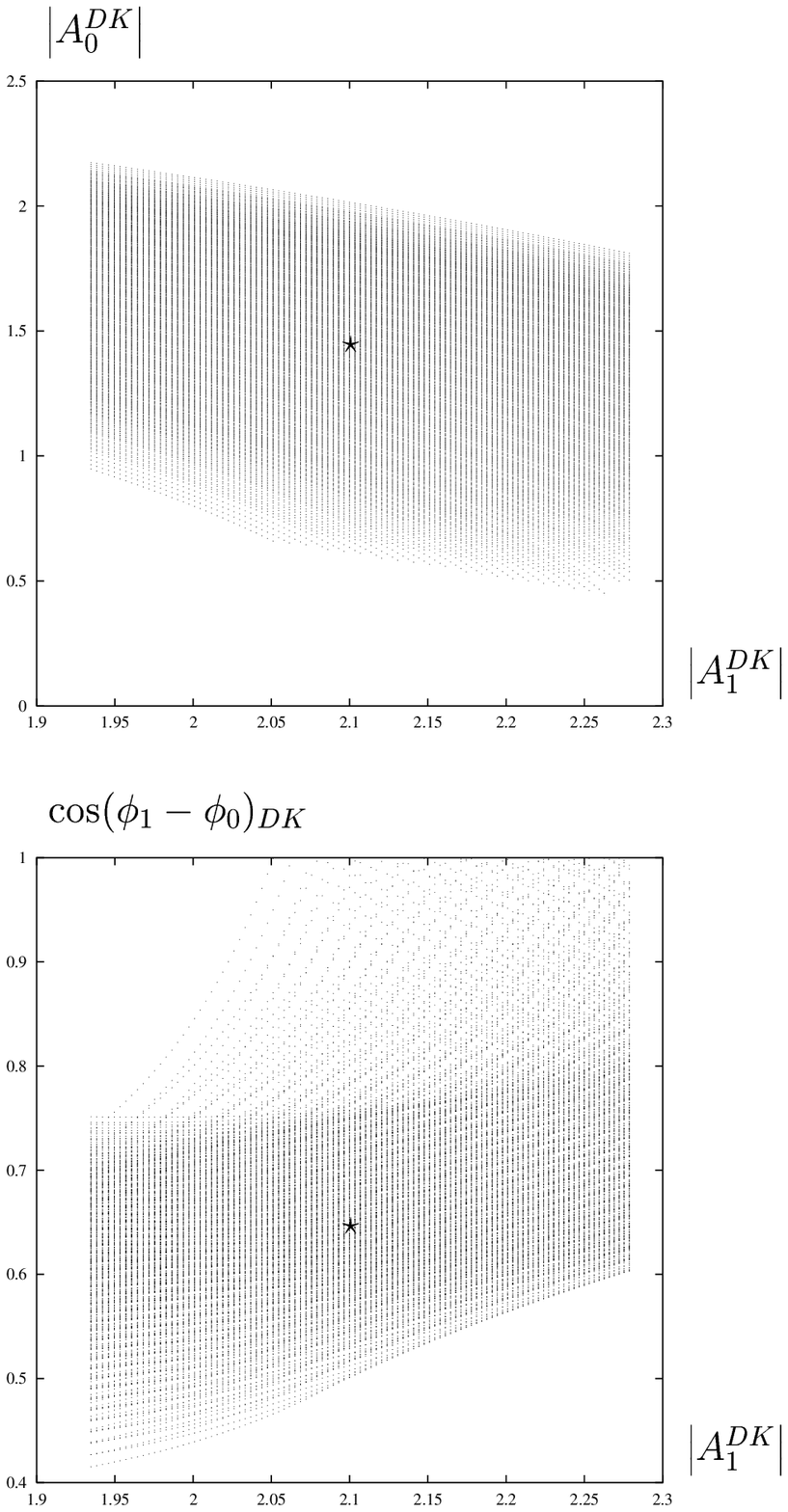,bbllx=1cm,bblly=5cm,bburx=16cm,bbury=22cm,%
width=13cm,height=15cm,angle=0,clip=90}
\vspace{-1.2cm}
\caption{The ranges of $|A^{DK}_1|$, $|A^{DK}_0|$ and
$\cos (\phi_1 -\phi_0)_{DK}$ allowed by current experimental data,
where $\star$ indicates the result obtained from the central 
values of ${\cal B}^{DK}_{+-}$, ${\cal B}^{DK}_{00}$ and 
${\cal B}^{DK}_{0-}$.}
\end{figure}

\begin{figure}[t]
\vspace{5cm}
\epsfig{file=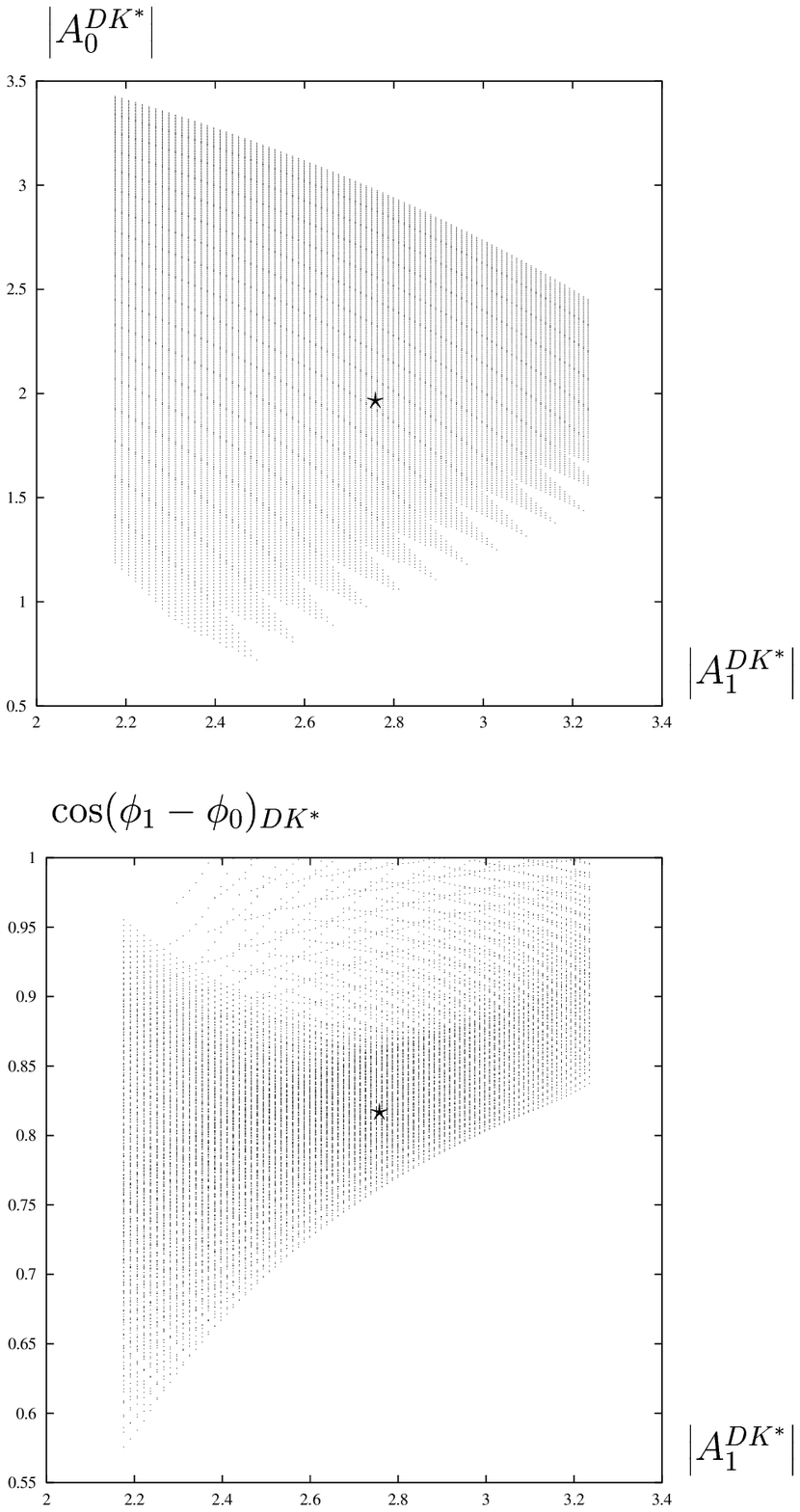,bbllx=1cm,bblly=5cm,bburx=16cm,bbury=22cm,%
width=13cm,height=15cm,angle=0,clip=90}
\vspace{-1.2cm}
\caption{The ranges of $|A^{DK^*}_1|$, $|A^{DK^*}_0|$ and
$\cos (\phi_1 -\phi_0)_{DK^*}$ allowed by current experimental data,
where $\star$ indicates the result obtained from the central 
values of ${\cal B}^{DK^*}_{+-}$, ${\cal B}^{DK^*}_{00}$ and 
${\cal B}^{DK^*}_{0-}$.}
\end{figure}

\begin{figure}[t]
\vspace{7cm}
\epsfig{file=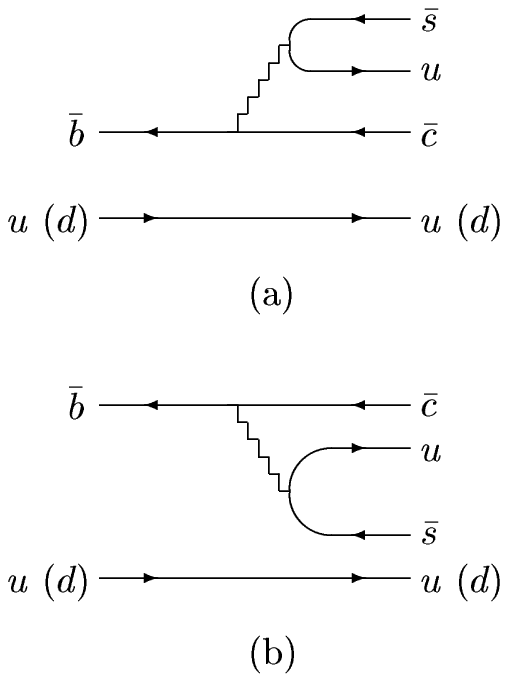,bbllx=4cm,bblly=6cm,bburx=16cm,bbury=18cm,%
width=15cm,height=16cm,angle=0,clip=90}
\vspace{-13cm}
\caption{The quark diagrams responsible for $B \rightarrow DK$ and 
$B \rightarrow DK^*$ decays: (a) the spectator; and (b) the
color-suppressed spectator.}
\end{figure}

\end{document}